\documentclass[11pt]{article}
\usepackage{amsmath,amssymb}
\usepackage{graphicx,color}
\usepackage{cite}
\usepackage[hypertex]{hyperref}
\newcommand{\be}{\begin{equation}}
\newcommand{\bea}{\begin{eqnarray}}
\newcommand{\eea}{\end{eqnarray}}
\newcommand{\ba}{\begin{array}}
\newcommand{\ea}{\end{array}}
\newcommand{\ee}{\end{equation}}
\newcommand{\nn}{\nonumber}\newcommand{\half}{\frac{1}{2}}
\newcommand{\A}{{\cal{A}}}

\newcommand{\M}{{\cal{M}}}
\newcommand{\C}{{\cal{C}}}
\newcommand{\T}{{\cal{T}}}
\newcommand{\CO}{{\cal{O}}}
\newcommand{\F}{{\cal{F}}}

\newcommand{\MC}{{\mathbb{C}}}

\newcommand{\p}{\partial}

\begin{document}

\begin{titlepage}
\thispagestyle{empty}

\begin{center}
\noindent{\Large \textbf{Holographic Entanglement Entropy\\ \vspace{0.2cm}for Excited States in Two Dimensional CFT
}}\\
\vspace{1cm}

%authors
\vspace{1cm}
Amin Faraji Astaneh$^{\rm a,b,}$\footnote{faraji@ipm.ir} and Amir Esmaeil Mosaffa $^{\rm b,}$\footnote{mosaffa@theory.ipm.ac.ir}

 \vspace*{0.25cm}
 \begin{quote}
 \center $^{\rm a}$\,\, {\sl Department of Physics, Sharif University of Technology,\\
P.O. Box 11365-9161, Tehran, Iran}

\center $^{\rm b}$\,\, {\sl School of Particles and Accelerators,\\ Institute for Research in Fundamental Sciences (IPM), \\
 P.O. Box 19395-5531, Tehran, Iran}

 \end{quote}

\end{center}
\vspace{2cm}
\begin{abstract}
We use holographic methods to study the entanglement entropy for 
excited states in a two dimensional conformal field theory. The
entangling area is a single interval and the excitations are produced
by in and out vertex operators with
given scaling dimensions. On the gravity side we provide the
excitations by turning on a scalar field with an appropriate mass.
The calculation amounts to using the gravitational background, with
a singular boundary, to find the one point function of the vertex
operators. The singular boundary is taken care of by introducing a
nontrivial UV regulator surface to calculate
gravitational partition functions. By means of 
holographic methods we reproduce the field theory results 
for primary excitations. 
\end{abstract}

\end{titlepage}

\newpage
\tableofcontents
\vspace{0.5 cm}
\section{Introduction}
There are many reasons why Entanglement Entropy (EE) has attracted a
lot of attention in the last few years. On the one hand it captures
information about the quantum structure of physical systems, to the
extent that it can be used as an order parameter to identify quantum
phases at zero temperature. On the other hand there are several
hints that this quantity can address outstanding questions in
quantum gravity.

Incidentally, EE first attracted considerable interest, 
\cite{'tHooft:1984re, Bombelli:1986rw, Srednicki:1993im}, when it was
shown to have similarities with the entropy of black holes through
the area law \cite{Bekenstein:1973ur, Hawking:1974sw}. This brought about the possibility that black
hole entropy might be explained as arising from entanglement between
degrees of freedom on the two sides of horizon.

Although these early hopes and speculations did not survive further
tests and expectations, in recent years new venues and frameworks
have been created to address and formulate old questions.
Holography, and gauge/gravity correspondence in a broader sense, has
been the subject of intense research activity in this regard 
\cite{Maldacena:1997re, Gubser:1998bc, Witten:1998qj}. It is
specially in this context that EE has proved to be very useful.

The proposal of \cite{Ryu:2006bv} 
(see also \cite{Ryu:2006ef,Nishioka:2009un,Takayanagi:2012kg} for reviews and references), gives a holographic prescription for
calculation of EE. This has enabled one to predict how this quantity
behaves in the strong coupling regime of certain quantum field
theories, a domain which is not accessible by the usual perturbative
methods. The results obtained this way suggest that the area law,
governing EE, remains generally valid in strongly coupled systems.
Furthermore, violations of the area law and/or the form of the
subleading contributions to EE contain various information about the
strongly coupled system, information such as whether the system has 
a Fermi surface, whether it is in a (de)confined
phase, whether it is a (non)local theory, etc.
(see \cite{Holzhey:1994we,Calabrese:2004eu, Calabrese:2005zw, Calabrese:2009qy,  
Eisert:2008ur, Huijse:2011ef, Fursaev:2006ih, Hubeny:2007xt, Klebanov:2007ws, Faraggi:2007fu,
Ogawa:2011bz, Shaghoulian:2011aa, Headrick:2010zt, deBoer:2011wk, 
Hung:2011xb, Casini:2011kv, Hung:2011ta, Hung:2011nu, Myers:2012ed, VanRaamsdonk:2009ar,
VanRaamsdonk:2010pw, Swingle:2012wq, Asplund:2011cq, Asplund:2012rg, Fischler:2012ca, Fischler:2012uv} for some related works, reviews and references).  

On the other hand, the above mentioned proposal gives a geometric
interpretation for EE. This may lead one into insights about the
mechanism of holography, how gravity comes into play, how additional
directions of space are generated in a gravitational description,
what regions of geometry are describing which experiments in field
theory and so on and so forth.

The idea of entanglement arises when one considers two or more
disjoint sets of degrees of freedom in a system. These degrees of
freedom can be far apart spatially and hence may be considered as
living in different regions of space. If there is a nonzero
entanglement between two such regions, local measurements and
observations in one region are affected by those in the other
instantaneously. The effect is obviously not carried by any
messenger, rather, it is the result of the quantum structure of the
state, that is, the way the overall quantum state of the system is
described in terms of the local quantum states in each region.

An observer that is confined to one of the regions will measure
effects which he cannot explain by the degrees of freedom accessible
to him and hence he will have some sort of $\it{ignorance}$ about
the accessible region. This ignorance or lack of information is
quantified by the EE.

Lack of information can also be caused by statistical distribution
of states in a system such as the case for thermal ensembles. In
such situations EE will no longer be a useful measure of quantum
entanglement and thus one usually studies this quantity when the
system is in a pure state. Most of the research on this subject has
focused on the case where this pure state is the ground state of the
theory. In this article we are interested in excited pure states.

As a first attempt one can consider primary excitations in a two
dimensional CFT and study the EE of a single interval. This problem
has been addressed in  \cite{Alcaraz:2011tn}, \cite{Berganza:2011mh} and
\cite{Mosaffa:2012mz} where three different methods have
been used for calculations. The first paper follows the approach of
Holzhey, Larsen and Wilczek \cite{Holzhey:1994we}, the second uses methods developed by 
Calabrese and Cardy \cite{Calabrese:2004eu} and the third one
applies techniques in symmetric orbifolding \cite{Lunin:2000yv}. In this work we will
address this problem by holography.

On the field theory side, one resorts to the replica trick for
calculation of the Renyi entropy which gives EE as a limit. This is
achieved by calculating the partition function of the theory on a
certain Riemann surface. The excitations are introduced by inserting
appropriate in and out vertex operators on this surface. One then
uses suitable conformal transformations to move over to a smooth
manifold on which the correlation function of the vertex operators
is calculated. The complexities of the Riemann surface, which
contain the nontrivial information of entanglement, are encoded in
the transformation to the smooth surface.

In this work we make a parallel line of calculations on the
gravitational side. We construct the space time out of the boundary
data and turn on certain fields with suitable boundary conditions to
account for excitations. This space will have a complicated Riemann
surface as the boundary. We will replace it with a smooth surface
and encode its complexities in a nontrivial UV regulator surface
near the boundary. The regulator will then be used to calculate the
gravitational partition function.

Our holographic calculations rely on the well developed techniques of
holographic renormalisation. We will use approximations to write down
the renormalized action on our nontrivial regulator surface. The approximations
we will use will turn out to be equivalent to the assumption
that on the field theory side some specific operators have been turned on.
The conformal properties of these operators are those for primary excitations.
From another point of view we may conclude that once we ``demand" our 
approximations to be exact, we have effectively given a bulk ``definition" 
for the field theory primary operators.

By the holographic calculation we therefore reproduce the field theory
results which have been obtained for $\it{primary}$
excitations of CFT.

In the following chapters we will first review EE in field theory
and its holographic description. In particular we will mention the
role of a regulator surface in the holographic calculation. We will
also briefly outline the field theory calculation of EE in presence
of primary excitations. We will then use gravity to address this
problem. We close with conclusions and discussions.

%%%%%%%%%%%%%%%%%%%%%%%%%%%%%%%%%%%%%%%%%%%%%%%%%%%%%%%%%%%%%%%%%%%%%%
%%%%%%%%%%%%%%%%%%%%%%%%%%%%%%%%%%%%%%%%%%%%%%%%%%%%%%%%%%%%%%%%%%%%%%

\section{Entanglement Entropy in Quantum Field Theory}
In this section we first give a short review of the basics of EE in
field theory and will then briefly outline the field theoretic
calculation of EE of a single interval in a two dimensional
conformal field theory in presence of primary excitations. The
latter will be addressed in a holographic setup in subsequent
sections.
%%%%%%%%%%%%%%%%%%%%%%%%%%%%%%%%%%%%%%%%%%%%%%%%%%%%%%%%%%%%%%%%%%%%%%
\subsection{Review of EE in QFT}
Consider a quantum mechanical system in a pure state  $ \vert \psi
\rangle $ and correspondingly with the density operator $ \hat{\rho}
=|\psi\rangle\langle{\psi}|$. Divide the system into two subsystems
$ A $ and $ B $ and define the $\it{reduced}$ density operator of
$A$ as $ \hat{\rho}_A = tr_B \hat{\rho} $. Alternatively one could
take the trace over the $A$ degrees of freedom and define $
\hat{\rho}_B$. Generically the reduced density operator will no
longer be pure and one can attribute entropy to it. The EE of
subsystem $A$ is defined as the \emph{Von-Neumann} entropy of this
operator
\begin{equation}
S_A =-tr_A \hat{\rho}_A \log \hat{\rho}_A.
\end{equation}
It is also possible to define and calculate EE in the context of
quantum field theory. A usual approach is to define a useful
mathematical quantity, called \emph{Renyi Entropy}, by the
\emph{Replica Trick} as
\begin{equation}
S_n[A]\equiv \frac{1}{1-n}\log tr_A \hat{\rho}_A^n.
\end{equation}
It is easy to see that EE can be obtained as $ S_A=
\lim_{n\rightarrow 1}S_n[A] $ whenever the limit exists. This
quantity can be represented in terms of a path integral by
considering $n$ copies of the world volume of the original theory,
$\mathcal{M}$, and glueing them along the entangling subspaces in a
cyclic order. This results in a space, which we denote by
$\mathcal{R}_n$, and which has singularities on the boundaries of
entangling subspaces. The path integral on $\mathcal{R}_n$ is
denoted by $Z_{\mathcal{R}_n}$ (or $Z_n$ in short) and is defined as
\be Z_{\mathcal{R}_n}=\int [d\varphi(x)]\ e^{-S[\varphi]}\ ,\ \ \
x\in\mathcal{R}_n\  , \ee where $n$ is the replica number and
$\varphi$ denotes dynamical fields collectively. One can then
calculate the Renyi entropy as
\begin{equation}
S_n = \frac{1}{1-n}\log \frac{Z_n}{Z_1^n},\label{Renyi}
\end{equation}
where $ Z_1 $ represents the partition function of the original
unreplicated theory.

Due to the complicated singular topology of the Riemann surface it
is usually very difficult to calculate this partition function
directly. One can go around this by transferring the geometric
complexities of the world volume into the geometry of target space.
That is, one considers the original nonsingular world volume but
instead introduces $n$ copies of the target space fields,
$\varphi_i\ (i=1,2,...,n)$, on that. Instead of gluing the world
volumes, one now restricts the fields to satisfy certain conditions
along the entangling subspaces
\begin{equation} Z_{res}=\int_{res}[d^n\varphi(x)]\
e^{-S[\varphi_1,...,\varphi_n]}\ ,\ \ \ x\in\mathcal{M}\ ,
\end{equation}
where the subscript $res$ stands for restrictions on
fields. Note that these restrictions replace the nontrivial geometry
of $\mathcal{R}_n$. One way to impose the restrictions is to insert
the so called $\it{twist \ operators}$ at the boundaries of
entangling subspaces and calculate an unrestricted integral
\begin{equation}
Z_{Twist}=\int_{unres}[d^n\varphi(x)]\
e^{-S[\varphi_1,...,\varphi_n]}\ \prod \sigma_k.....\ ,\ \ \
x\in\mathcal{M}\ , \ee where $\sigma_k$ are the twist operators that
open and close the branch cuts along the entangling subspaces and
enforce the restrictions through their Operator Product Expansion
(OPE) with fields. An alternative way of imposing the restrictions
is to move over to the covering space of the fields, denoted by
$\mathcal{M}_C$, with a suitable coordinate transformation and
perform the calculations on this manifold \be\label{coveringspace}
Z_{\mathcal{M}_C}=\int [d\varphi(x)]\ e^{-S[\varphi]}\ ,\ \ \
x\in\mathcal{M}_C\ .
\end{equation}
On the covering space the restrictions on fields in the integration
are taken care of by the geometry of $\mathcal{M}_C$. This space is
now a smooth manifold and the complexities of the Riemann surface
are encoded in the transformation to $\mathcal{M}_C$\footnote{Look at transformations
(\ref{cyltop}) and (\ref{conformal map}) as examples.}.
%%%%%%%%%%%%%%%%%%%%%%%%%%%%%%%%%%%%%%%%%%%%%%%%%%%%%%%%%%%%%%%%%%%%%
\subsection{Entanglement entropy for primary excitations in 2-d CFT}\label{excitationft}
In this section we briefly outline calculation of EE for a single
interval in a two dimensional CFT in presence of primary
excitations. For details see \cite{Alcaraz:2011tn, Berganza:2011mh, Mosaffa:2012mz}.

We start with a CFT on a cylinder, parametrized by  $(x,\bar{x})$, where
$x=\sigma+i\tau$ and $\sigma\in[0,2\pi]\ ,
\tau\in(-\infty,+\infty)$. The entangling surface, $\mathcal{A}$, is the interval
$x\in[a,b]$.
To prepare the system in a highest weight
excited state, we generate the asymptotic \emph{incoming} and
\emph{outgoing} states by inserting the corresponding operators $
\CO(x,\bar{x}) $ and  $ \CO^\dagger(x,\bar{x}) $ at infinite past
and future time, ($ \tau\rightarrow \mp\infty)$, respectively.
The quantity of interest is
\begin{equation}
tr\rho^n_\CO\equiv\frac{Z_n}{Z_1^n}\lim_{x,\bar{x}\rightarrow -i\infty}
\frac{\langle \displaystyle \prod_{k=0}^{n-1}\CO_k(x,\bar{x})
\CO_k^\dagger(-x,-\bar{x}) \rangle_{\mathcal{R}_n}}{\langle
\CO(x,\bar{x})\CO^\dagger(-x,-\bar{x})\rangle^n_{x}}\ ,\label{corrcylinder}
\end{equation}
where the subscripts $\mathcal{R}_n$ and $x$ denote correlation on
the Riemann surface with $n$ cylinders and that on a single cylinder
respectively. The subscript $k$ labels the cylinder on which we have inserted the operators.
We can now move over to the smooth $z$ plane by
\begin{equation}
z=\left[\frac{e^{-ix}-e^{-ia}}{e^{-ix}-e^{-ib}}\right]^\frac{1}{n}\ ,\label{cyltop}
\end{equation}
and use the transformation properties of the primary operators to
calculate the relevant factors to arrive at
\begin{equation}
tr\rho^n_\CO =\frac{Z_n}{Z_1^n}\T\,\frac{\langle \displaystyle
\prod_{k=0}^{n-1}\CO(z_k,\bar{z}_k)\CO^\dagger(z'_k,\bar{z}'_k)
\rangle_{\MC}}{\langle \CO(\zeta,\bar{\zeta})
\CO^\dagger(\zeta',\bar{\zeta}')\rangle^n_{\MC}}\ ,\label{corrplane}
\end{equation}
where the subscribe $ \MC $ stands for the complex plane and $ \T $
is coming from the conformal transformation
(\ref{cyltop}). Note that the points $x=(a,b)$ map to $z=(0,\infty)$.
In addition, $ z_k $ and $ z'_k $ are representing
the insertion points, $x=-i\infty$ and $x=i\infty$ respectively,  which lie on the
circumference of a circle with unit radius on the complex plane $ \MC$
\begin{equation}
\zeta=e^{i\pi\theta}\ ,\quad \zeta'=e^{-i\pi\theta}\ ,\quad z_k=e^{\frac{i\pi}{n}(\theta +2k)}\quad , \quad
z'_k=e^{\frac{i\pi}{n}(-\theta+2k)}\quad , \quad k=0, 1,\cdots
,n-1\quad,\quad\label{insertions}
\end{equation}
where $ \theta\equiv\frac{b-a}{2\pi}\equiv\frac{l}{2\pi}$ .
The nontrivial information
will be included in the transformation factor $\T$ and the
theory dependent correlation functions on the plane. The change in the EE
coming from the insertions are calculable from the quantity
\begin{equation}
\F_\CO^{(n)}\equiv \frac{tr \rho^n_\CO}{tr \rho^n}.
\end{equation}
%%%%%%%%%%%%%%%%%%%%%%%%%%%%%%%%%%%%%%%%%%%%%%%%%%%%%%%%%%%%%%%%%%%%%%%%
%%%%%%%%%%%%%%%%%%%%%%%%%%%%%%%%%%%%%%%%%%%%%%%%%%%%%%%%%%%%%%%%%%%%%%%%
\section{Holographic Entanglement Entropy}
In this section we will first state the \emph{Ryu-Takayanagi's}
proposal for a holographic description of EE. We then focus on two
dimensional field theory living on a singular Riemann surface and
review how its holographic space-time can be constructed. The latter
will be used for a direct calculation of partition function.
%%%%%%%%%%%%%%%%%%%%%%%%%%%%%%%%%%%%%%%%%%%%%%%%%%%%%%%%%%%%%%%%%%%%%%%%
\subsection{\emph{Ryu-Takayanagi's} proposal}
One interesting approach towards the calculation of the EE in a CFT,
by applying the AdS/CFT correspondence is
proposed in \cite{Ryu:2006bv}.\\
According to the \emph{Ryu-Takayanagi's} proposal, one can find the
EE for subregion $ \A $ in $CFT_d$ holographically by using the
following area law relation
\begin{equation}
S_\A=min_{\gamma_\A}\left[\frac{Area(\gamma_\A)}{4G_N^{d+1}}\right],
\end{equation}
where $ G_N $ is the Newton constant and $ \gamma_\A $ is a
codimmension two surface in the bulk geometry whose boundary
coincides with the boundary of subregion $ \A $, $ \p\gamma_\A = \p \A $.\\
As a simple example, where subregion $ \A $ is a single interval in a
(1+1)-dim CFT, we must compute the length of the geodesic line in $
AdS_3 $ space to find the EE. Then it is straightforward to show
that this entropy exactly yields the known 2-dim CFT result.
\paragraph{}
An alternative approach for calculating the EE holographically which
is more convenient for our purpose, is based on a direct holographic
calculation to find the CFT partition function
from the classical gravity one.\\
The following subsection provides a brief introduction to this
method for a one dimensional single interval problem.
%%%%%%%%%%%%%%%%%%%%%%%%%%%%%%%%%%%%%%%%%%%%%%%%%%%%%%%%%%%%%%%%%%%%%%%
\subsection{Direct holographic calculation}\label{directholo}
In the remainder of this section, following \cite{Hung:2011nu}, we
review the direct calculation of Renyi entropy for the special case
of a single interval in two dimensional CFT by holography.

The system lives
on a plane, parametrized by $(w,\bar{w})$ and the entangling space is
the interval $w\in[u,v]$.
This is related to the cylinder of section (\ref{excitationft}) by
\be
w=e^{-ix}\ , \quad u=e^{-ia}\ ,\quad v=e^{-ib}\ . \label{cyltopl}
\ee
The starting point is the theory on $\mathcal{R}_n$ that
has been produced through
replication and consists of $n$ copies of the $w$-plane, glued
together along the entangling space in a cyclic order. This produces
a concentration of curvature at the boundaries of
entangling subspaces. To give a
holographic description of the problem we thus need to find a three
dimensional space-time that has $\mathcal{R}_n$ as the boundary.
This can be achieved by the ``holographic reconstruction of
spacetime" \cite{Skenderis:1999nb}, \cite{deHaro:2000xn}.

It is more convenient to work in the \emph{Fefferman-Graham} (FG)
gauge, mainly because all of the required nontrivial regularizations
during the calculation of the EE take very simple forms in this
gauge. The metric on the boundary is found through the map
\begin{equation}
z=\left(\frac{w-u}{w-v}\right)^\frac{1}{n}\ ,\label{conformal map}
\end{equation}
that takes $\mathcal{R}_n$
to the flat $z$-plane.
One finds that the metric is conformally flat,
\be
ds^2=\C (w,\bar{w},u,v) \ dw d\bar{w}\ , \label{conffactor}
\ee
with the expected
curvature at singular points
found by $R_0^w\sim\nabla_w^2\log\C$. Through trace anomaly, this in
turn will be related to the expectation value of the boundary
energy-momentum tensor, $\langle
T_{ww}\rangle\equiv-c/(24\pi)\mathcal{T}_{ww}$, where $c$ is the
central charge of the CFT.

One can now use a suitable regularization to calculate the
gravitational on-shell action which consists of the Einstein-Hilbert
part and the boundary contributions. Putting everything together one
arrives at
\begin{equation}
S_{grav}=-\frac{c}{12\pi}\int dw d\bar{w}\sqrt{g_{(0)}}
\left( \frac{1}{4}R_0\log \rho^* +
\half \sqrt{\vert \T_{ww}\vert^2}+\frac{3}{8}R_0\right),
\end{equation}
where $ \rho^* $ stands for the upper limit of the integral in the
radial direction of space (see \cite{Hung:2011nu} for details). A
careful calculation gives
\begin{equation}
S_{grav}=\frac{c}{6}\left(n-\frac{1}{n}\right)
\log (\frac{l}{\delta}),\label{final action}
\end{equation}
which, through eq.(\ref{Renyi}), results in the following expression
for the Renyi entropy
\begin{equation}
S_n=\frac{1}{n-1} S_{grav}=\frac{c}{6}
\left( 1+ \frac{1}{n} \right) \log (\frac{l}{\delta}).
\end{equation}
In the last two expressions $ l=v-u $ and $\delta$ is the short
distance regulator around the singular points.

%%%%%%%%%%%%%%%%%%%%%%%%%%%%%%%%%%%%%%%%%%%%%%%%%%%%%%%%%%%%%%%%%%%%%%

\subsection{The role of the UV regulator}\label{uvregulator}
The results obtained above, as noticed in \cite{Hung:2011nu}, may be attributed to a nontrivial
regularization in the bulk integrations. The regulator surface
extends the notion of the field theory short distance cut-off into
the bulk by cutting-off the radial integration near the boundary.
%$ \vr=\vr_{min}=\delta^2 $.
It also keeps the integrand away from the singularities at $ w=u $
and $ w=v $. The (FG) gauged was used to simplify this procedure.

Alternatively, one can work in the usual poincare coordinates
\begin{equation}
ds^2=\frac{1}{r^2}(dr^2 + dz d\bar{z}),\label{poincare}
\end{equation}
and instead deform the usual \emph{constant radius} regulator
surface appropriately to obtain the former results. This can be seen
by applying a suitable transformation between FG and Poincare
coordinates, as follows \cite{Krasnov:2001cu}
\begin{equation}
r=\frac{\rho^\half e^{-\psi}}{1+\rho e^{-2\psi}\vert
\p_y\psi\vert^2}\quad , \quad z=y+\p_{\bar{y}}\psi\frac{\rho
e^{-2\psi}}{1+\rho e^{-2\psi}\vert \p_y\psi\vert^2}\ ,\label{FGtoP}
\end{equation}
where
\begin{equation}
y\equiv\left(\frac{w-u}{w-v}\right)^\frac{1}{n},\ \ \ \ e^{\psi}=\frac{n}{l}\ \vert w-u\vert^{(1-1/n)}\ \vert w-v\vert^{(1+1/n)}=\vert \frac{dw}{dy}\vert^2\ .
\end{equation}
It is obvious that for $\rho\sim\epsilon^2\rightarrow 0$, this is the
desired conformal map between $w$-plane and its universal cover. The
metric (\ref{poincare}) in this new coordinate reads as
\begin{equation}
ds^2=\frac{d\rho^2}{4\rho^2}+\frac{g_{ij}(\rho,w,\bar{w})}{\rho}
dx^i dx^j\ , \label{FGcoord}
\end{equation}
where 
\be
i,j=1,2\ ,\quad g_{ij}(\rho,w,\bar{w})={g_{(0)}}_{ij}+\rho {g_{(2)}}_{ij}+\cdots\ ,\quad {g_{(0)}}_{ij}dx^idx^j=dwd\bar{w}\ .
\ee
One can see that the auxiliary field $\psi(z,
\bar{z})$ is nothing but the \emph{Liouville field}.

Now consider the following regulator surface
\begin{equation}
r=\epsilon\  e^{{-\varphi}(z,\bar{z})} \quad , \quad e^{\varphi(z,\bar{z})}=n l
\frac{\vert z \vert^{n-1}}{\vert z^n -1 \vert^2}\ .\label{regulatorsurf}
\end{equation}
Note that in the limit of $\epsilon\ll1$, $\psi$ reduces to $\varphi$ and hence $e^{2\varphi}=\vert dw/dz\vert^2$.
It can be shown that the induced metric on this surface has the same
singular conformal factor $\C$ in (\ref{conffactor}).
Using this
regulator, we find that the gravitational on-shell action in the
Ponicare coordinate will only depend on the auxiliary scalar field $
\varphi(z,\bar{z}) $ through (see \cite{Hung:2011nu} and \cite{Krasnov:2001cu})
\begin{equation}
S_{grav}=-\frac{c}{48\pi}\int dz d\bar{z}
(\p \varphi \bar{\p}\varphi - 4 \p\bar{\p}\varphi ).\label{gaction2}
\end{equation}
Substituting the explicit form of  $ \varphi(z,\bar{z}) $ in this
expression one can reproduce the former results in the (FG) gauge,
eq.(\ref{final action}). We will use this approach in our
holographic calculations of the following section.
%%%%%%%%%%%%%%%%%%%%%%%%%%%%%%%%%%%%%%%%%%%%%%%%%%%%%%%%%%%%%%%%%%%%%%%%%%%%
%%%%%%%%%%%%%%%%%%%%%%%%%%%%%%%%%%%%%%%%%%%%%%%%%%%%%%%%%%%%%%%%%%%%%%%%%%%%
\section{Holographic calculation of the entanglement entropy for excited states}
In this section we present our main calculations and results of this paper.
As explained before, we are interested in a general CFT on a circle which has been
excited by primary operators. We would like to calculate the EE of a single interval
in presence of these excitations. In the following we present a holographic
description of this problem which has already been addressed in a field theoretic 
setup and which has been shortly reviewed in section (\ref{excitationft}).
%%%%%%%%%%%%%%%%%%%%%%%%%%%%%%%%%%%%%%%%%%%%%%%%%%%%%%%%%%%%%%%%%%%%%%%%%%%%
\subsection{Holographic Setup}
Start with the CFT on the infinite $x$-cylinder of section (\ref{excitationft}), with insertions 
at infinite past and future time to account for excitations. 
The replicated theory consists of $n$ such cylinders, 
glued cyclically along the entangling interval, 
denoted as before by $\mathcal{R}_n$. This also
produces $n$ copies of operator insertions for in and out states each.

The objective is to calculate the path integral on $\mathcal{R}_n$
by holography. This will give us the Renyi, and subsequently, the
entanglement entropy. One can try to find a gravitational background
which has this replicated cylinder at the boundary. The operator
insertions will amount to turning on appropriate fields in the 
background. The gravitational partition function thus found
will be, according to AdS/CFT, our desired result.
This program is in principle possible through
the ``holographic reconstruction of space-time" 
and would be a generalization of section (\ref{directholo})
to the case with excitations.
 
We will choose, however, to take a different route and find the
gravitational analogue of equation (\ref{coveringspace}). That is,
we prefer to deal with the CFT on the smooth manifold, $\M_C$, and
worry about the complexities of the replicated cylinder only 
through the transformation $\mathcal{R}_n\rightarrow\M_C$.
This choice will give us the advantage that we should now
look for, and work with, a gravitational background that 
will have a smooth boundary rather 
than a singular one. The nontrivial information of $\mathcal{R}_n$,
and accordingly those in the transformation to $\M_C$, 
will be encoded in the regularization scheme that we choose to calculate
the gravitational partition function along the lines of section (\ref{uvregulator}).

The setup will thus be as follows; the flat complex $z$-plane, also denoted by $\mathbb{C}$,  is our $\M_C$ 
and sits on the boundary with insertion points given by (\ref{insertions}). 
The transformation $\mathcal{R}_n\rightarrow\M_C$ is nothing 
but (\ref{cyltop}). 
The gravitational background will thus be pure $AdS_3$ on which we 
will turn on a scalar field with suitable scaling properties 
and boundary conditions to account for the excitations. We will then
use the map (\ref{cyltop}) to find the \emph{Liouville} field that 
determines the shape of our regulator surface. We use this regulator
to find the renormalized on-shell gravitational and matter actions.
The standard recipe of AdS/CFT will then give us the exact one point 
function of the boundary operator and the desired correlation functions.

As a matter of simplifications, we map the replicated $x$-cylinders to 
replicated $w$-planes of section (\ref{directholo}) by the map of
(\ref{cyltopl}).  In the case of possible confusion, we differentiate between the two
by $x$ and $w$ superscripts  respectively.
The link between the replicated $w$-planes with
the $z$-plane  will be provided by (\ref{conformal map}). As a final
step we will map the $z$-plane to cylinder again.
Schematically
\be
\mathcal{R}_n^{(x)}\xrightarrow{w=e^{-ix}} \mathcal{R}_n^{(w)}\xrightarrow{eq.\ (\ref{conformal map})} z{\rm{-plane}}
\xrightarrow{s=-i\ln{z}}s{\rm{-cylinder}}\ ,
\ee
such that in the end we have a map between a replicated cylinder to a smooth one. 
We have introduced the intermediate manifolds, i.e., $\mathcal{R}_n^{(w)}$ and $z$-plane (or $\mathbb{C}$)
to simplify the holographic calculations and it is in fact this step $\mathcal{R}_n^{(w)}\rightarrow\mathbb{C}$
that we will take by holography in the following section.

The quantities we will find with the nontrivial regularization will
be in terms of those obtained with the standard simple cutoff regulator.
The former quantities are identified with CFT correlators on 
$\mathcal{R}_n$ and the latter with those on $\M_C$ which is
simply the complex plane $\MC$. This will in fact be the 
holographic analogue of going from equation (\ref{corrcylinder}) to
(\ref{corrplane}) in section (\ref{excitationft}). The role
of the linking equation (\ref{cyltop}) in field theory will be played
by the \emph{Liouville} field and the regulator surface in gravity.
In the next section we will show how this works.
%%%%%%%%%%%%%%%%%%%%%%%%%%%%%%%%%%%%%%%%%%%%%%%%%%%%%%%%%%%%%%%%%%%%%%
\subsection{Calculations}
We use the standard method of ``holographic renormalization"  \cite{deHaro:2000xn} 
(see also \cite{Skenderis:2002wp} for review and references) to extract physical quantities of the 
boundary CFT out of the bulk.
The quantity we should compute consists of two parts
\be
S=S_g+S_m\ ,\nn
\ee
where $S_g$ and $S_m$ refer to the gravitational and matter actions respectively. 
We will use two sets of coordinates and metrics for the space
which are collectively denoted by $(x^\mu, G_{\mu\nu})$.  
One is the usual Ponicare coordinates $(r,z,\bar{z})$ with the metric (\ref{poincare}) and the other one is
the FG coordinates $(\rho,w,\bar{w})$ with the metric (\ref{FGcoord}). These two are related by
(\ref{FGtoP}).
\newpage
\vspace{0.2cm}

\underline{\emph{Regulators}}

\vspace{0.3cm}

Physical quantities are obtained upon calculating the renormalized actions in bulk. This
in turn requires a regularization scheme.  A surface at a constant Poincare radius, $r=\epsilon$,
which we will denote by $\mathcal{M}_0$ will be used to calculate the physical quantities of the CFT 
on the $z$-plane, $\mathbb{C}$.
The regulator surface of (\ref{regulatorsurf}), 
denoted by $\mathcal{M}$, is attributed to the CFT on $\mathcal{R}_n$. A closely related surface which will play a crucial role for us is obtained as
a constant radius surface in the FG coordinates, $\rho=\epsilon^2$, and which we will denote by $\mathcal{M}'$.
To summarise our notation
\be
r=\epsilon\equiv\mathcal{M}_0\ ,\ \ \ \ r=\epsilon e^{-\varphi}\equiv\mathcal{M}\, \ \ \ \ \rho=\epsilon^2\equiv\mathcal{M}'\ .
\ee
The objective is to calculate the renormalized action on $\mathcal{M}$ and this can be achieved by going 
through the complete process of
holographic renormalisation. This is in principle doable but recall that in the limit of $\epsilon\ll1$
the regulators $\mathcal{M}$ and $\mathcal{M}'$ are equivalent and will therefore result in identical physical quantities.
The fact that $\mathcal{M}'$ has a simple form in the FG coordinates facilitates the regularization process. 
We will take advantage of this fact and use the well studied renormalized action on $\mathcal{M}'$.

The surface $\mathcal{M}$, on the other hand, has the advantage that as compared to $\mathcal{M}'$ has a
simpler description in the Poincare coordinates. This will enable us to write down the fields on this
surface in terms of those in the Poincare coordinates in a simple way which will eventually lead to a 
clear relation between physical quantities on $\mathcal{M}$ and those on $\mathcal{M}_0$. This, in turn,
will be used to describe the field theory on $\mathcal{R}_n$ in terms of the theory on $\mathbb{C}$.
We will expand on the details in the course of the calculations.

The plan therefore will be as the following. We will write the renormalized action on $\mathcal{M}_0$ and extract from it
the quantities attributed to the theory on $\mathbb{C}$. We will then find the renormalized action on $\mathcal{M}'$
and use perturbative methods to approximate the renormalized action on $\mathcal{M}$. The latter will be
used to extract physical objects of the theory on $\mathcal{R}_n$ which are related to those on $\mathbb{C}$
in a simple way. 

\vspace{0.2cm}

\underline{\emph{Action}}

\vspace{0.3cm}

The gravitational action $S_g$ using the nontrivial regulator $\mathcal{M}$ 
has already been studied in  \cite{Hung:2011nu} and their
results hold here. We will thus focus on the scalar part.
The action for a free scalar in bulk is given by
\begin{equation}
S_{m}=\half \int d^3 x\sqrt{G}(G^{\mu\nu}
\p_\mu\Phi\p_\nu\Phi + m^2\Phi^2)\ .
\end{equation}
The scalar field in the Poincare and FG
coordinates are denoted by $\Phi$ and $\Phi'$  and are related by
\be
\Phi(r,z,\bar{z})=\Phi'(\rho,w,\bar{w})\ .\label{scalars}
\ee
The asymptotic behaviour of the scalar in the Poincare coordinate is given by
\begin{equation}
\Phi(r,z,\bar{z})= r^{2-\Delta}\,\phi(r,z,\bar{z})\ ,\ \ \ \phi(r,z,\bar{z})=\phi_0(z,\bar{z}) + r^2 \phi_2(z,\bar{z}) + \cdots\label{scalarexp}
\end{equation}
where, dots stand for terms with higher powers as well as logarithmic terms in $r$.
Equivalently we can expand the scalar in the FG coordinates as
\be
\Phi'(\rho,w,\bar{w})=\rho^{\frac{2-\Delta}{2}}\phi'(\rho,w,\bar{w})\ ,\ \ \ \phi'(\rho,w,\bar{w})=\phi'_0(w,\bar{w})+\rho\phi'_2(w,\bar{w})+\cdots
\ee
Substituting any of the two expressions in the scalar field equation
results in the familiar relation between the mass of the scalar
field and the conformal dimension of the dual operator, $\Delta$, as $m^2=\Delta(2-\Delta) $.

Let us first take the regulator surface to be $\mathcal{M}_0$. The \emph{regularized} matter action
is obtained when the integration region is bounded by this surface
\begin{align}
&S_{m,reg}=\half \int_{r\geqslant\epsilon} d^3x
\sqrt{G}(G^{\mu\nu}\p_\mu\Phi\p_\nu\Phi +m^2\Phi^2)=\nonumber\\
&-\half\int_{r=\epsilon}\sqrt{G}\,dzd\bar{z}\ G^{rr}\Phi\p_{r}\Phi=-\half\int_{r=\epsilon}\sqrt{\gamma} dzd\bar{z}\ r\Phi\partial_r\Phi\ ,\label{regact}
\end{align}
where $\gamma_{ij}=1/r^2\delta_{ij}$ is the induced metric on a constant $r$ surface.
Also note that  we have used the equations of motion to drop the bulk action.
The \emph{subtracted} action is defined upon the addition of suitable counterterms to the regularized action\footnote{For the details of this
procedure see \cite{deHaro:2000xn}.}
\be
S_{sub}(\Phi,\gamma)_{\mathcal{M}_0}= \int_{r=\epsilon}\sqrt{\gamma}dzd\bar{z}\ \Phi[-\frac{r}{2}\partial_r\Phi+\frac{2-\Delta}{2}\Phi+
\frac{1}{2(\Delta-2)}\Box_{\gamma}\Phi]\ ,\label{subactPoinc}
\ee
where $\Box_{\gamma}$ stands for the Laplacian operator made by the induced metric and 
we have dropped some higher derivative terms in the counterterm part. 
Similarly on the regulator surface $\mathcal{M}'$ we find\footnote{We avoid singular points on the surface
by limiting the integration to the regular parts. This will have no effect for the matter action and we thus do not mention it
in the integration limits. As for the gravitational action the singular points play a crucial role (see \cite{Hung:2011nu}).}
\be
S_{sub}(\Phi',\gamma')_{\mathcal{M}'}=\int_{\rho=\epsilon^2}\sqrt{\gamma'}dwd\bar{w}\ \Phi'[-\rho\partial_{\rho}\Phi'+\frac{2-\Delta}{2}\Phi'+
\frac{1}{2(\Delta-2)}\Box_{\gamma'}\Phi']\ ,\label{subactFG}
\ee
where $\gamma'_{ij}=g_{ij}(\rho,w,\bar{w})/\rho$ and 
we are using ``primed" quantities to emphasise that on a constant $\rho$ surface 
everything is naturally written in terms of the FG coordinates and $\Phi'$.

\vspace{0.3cm}

\underline{\emph{$\mathcal{M}'$ vs. $\mathcal{M}$}}

\vspace{0.3cm}
We now wish to use (\ref{subactFG}) to calculate the renormalized action on $\mathcal{M}$. In order to do this,
we should expand (\ref{subactFG}) in the limit of $\epsilon\ll1$. Note that in this limit the transformation (\ref{FGtoP}) reduces to
\be
r=\rho^{1/2}e^{-\varphi}\ ,\ \ \ \ z=\left(\frac{w-u}{w-v}\right)^{1/n}\ .\label{FGtoPapprox}
\ee
The relation between $z$ and $w$ now has no $\rho$ dependence and 
describes the conformal transformation from $\mathcal{R}_n$ to $\mathbb{C}$.
Also $\rho=\epsilon^2$, or equivalently $r=\epsilon e^{-\varphi}$, is now describing the regulator surface of $\mathcal{M}$.
On this surface we can approximate the scalar field $\Phi'$ by 
$\tilde{\Phi}(\rho,w,\bar{w})=\Phi(r,z,\bar{z})$. We are using a $\tilde{\Phi}$ instead of $\Phi'$ and
call it an approximation because we are using the approximate transformation (\ref{FGtoPapprox}). Since the
$\epsilon$ dependence is now entering only through the relation between $r$ and $\rho$, the coefficients
of an $\epsilon$ expansion can be easily extracted and compared in the two coordinates
\be
\tilde{\Phi}(\epsilon^2,w,\bar{w})=\Phi(\epsilon e^{-\varphi},z,\bar{z})\ \Rightarrow
\tilde{\phi}_{2n}=(e^{-\varphi})^{2n+2-\Delta}\ \phi_{2n}\ ,\label{expansions}
\ee
where
\be
\tilde{\Phi}(\rho,w,\bar{w})=\rho^{\frac{2-\Delta}{2}}\tilde{\phi}(\rho,w,\bar{w})\ ,\ \ \ \tilde{\phi}(\rho,w,\bar{w})=
\tilde{\phi}_0(w,\bar{w})+\rho\tilde{\phi}_2(w,\bar{w})+\cdots
\ee
We can now write down the approximate form of (\ref{subactFG}) by replacing $\Phi'$ with $\tilde{\Phi}$ and remembering that 
the transformation to the Poincare coordinates is now given by (\ref{FGtoPapprox}). We denote the resulting
approximate action by $\Tilde{S}_{sub}(\tilde{\Phi},\gamma')_{\mathcal{M}}$. Again the tilde is there to remind us of
the approximation.

The action $\tilde{S}$ thus found is a good approximation to the subtracted action on the regulator surface $\mathcal{M}$
for small $\epsilon$ and for a sufficiently slowly varying $\varphi$. In principle there will be corrections in derivatives of
$\varphi$ such that physical quantities constructed from $\tilde{S}$ yield finite values. 
We ignore these corrections
in the following and will comment on the validity of our assumption later.

Note that the action on $\mathcal{M}'$ is an exact one but a simple relation like (\ref{expansions}) is missing
in that case. Conversely, such a relation exists for $\mathcal{M}$ but instead the action we found is approximate.

\vspace{0.3cm}

\underline{\emph{Interpretation of $\tilde{S}$}}

\vspace{0.3cm}
The action $\tilde{S}_{\mathcal{M}}$ can be interpreted as the 
familiar holographic description of a conformal transformation 
on the boundary (deformed) CFT. Instead of the usual constant scaling in the radial AdS direction, we have 
been using a local scaling factor which has been forced on us by the geometry of $\mathcal{R}_n$.
Let us see how this works.

Recall that conformal transformation is a sequence of a Weyl scaling of the metric followed by a coordinate transformation. 
Once we rewrite $\tilde{S}_{\mathcal{M}}$ in terms of quantities and fields which are naturally defined in the 
Poincare coordinates we will have described the Weyl scaling. For this purpose we may first easily check that
for small $\epsilon$ and on $\mathcal{M}$
\be
\sqrt{\gamma'}dwd\bar{w}=\sqrt{\tilde{\gamma}}dzd\bar{z} ,\ 
\rho\partial_{\rho}\tilde{\Phi}=\frac{r}{2}\partial_r\Phi\ ,\ 
\Box_{\gamma'}\tilde{\Phi}=\Box_{\tilde{\gamma}}\Phi\ ,
\ee
where $\tilde{\gamma}_{ij}=\gamma_{ij}e^{2\varphi}$. Plugging these in $\tilde{S}$ we find that 
\be
\tilde{S}_{sub}(\tilde{\Phi},\gamma')_{\mathcal{M}}=\int_{\mathcal{M}}\sqrt{\tilde{\gamma}}dzd\bar{z}\ \Phi[-\frac{r}{2}\partial_r\Phi+
\frac{2-\Delta}{2}\Phi+\frac{1}{2(\Delta-2)}\Box_{\tilde{\gamma}}\Phi]=S_{sub}(\Phi,\tilde{\gamma})_{\mathcal{M}}\ .\label{subacttilde}
\ee
It can be seen that $S_{sub}(\Phi,\tilde{\gamma})_{\mathcal{M}}$, thus obtained, is nothing but $S_{sub}(\Phi,\tilde{\gamma})_{\mathcal{M}_0}$
with a scaled metric and of course with a different argument for $\Phi$. To complete the conformal transformation we now 
make a coordinate transformation that cancels out the Weyl factor of the metric.
This is of course the transformation $(z,\bar{z})\rightarrow(w,\bar{w})$ in (\ref{FGtoPapprox}) which takes us back to 
$S_{sub}(\Phi,\tilde{\gamma})_{\mathcal{M}}$.
\vspace{0.3cm}

\underline{\emph{One point function}}
\vspace{0.3cm}

We can now extract the exact one point function of the field theory operator that is dual to the bulk scalar field. As for the
theory on $\mathbb{C}$, this is obtained as
\footnote{We should note that the variations of the regularized 
parts of the subtracted actions are taken from the regularized bulk actions before imposing the equations of motion, i.e., from the first
line in (\ref{regact}).}
\be
\langle\mathcal{O}(z,\bar{z})\rangle_{\mathbb{C}}=\lim_{\substack{\epsilon\rightarrow0}}
\left[\frac{1}{r^{\Delta}\sqrt{\gamma}}\frac{\delta S_{sub}(\Phi,\gamma)}{\delta\Phi}\right]_{\mathcal{M}_0}=(2-2\Delta)\phi_{(2\Delta-2)}\ .
\ee
The one point function for the CFT on $\mathcal{R}_n$ can be obtained either from $S_{sub}(\Phi',\gamma')_{\mathcal{M}'}$ in 
(\ref{subactFG}) or, equivalently, from the subtracted action on $\mathcal{M}$. Using the former we get
\be
\langle\mathcal{O'}(w,\bar{w})\rangle_{\mathcal{R}_n}=\lim_{\substack{\epsilon\rightarrow0}}
\left[\frac{1}{\rho^{\Delta/2}\sqrt{\gamma'}}\frac{\delta S_{sub}(\Phi',\gamma')}{\delta\Phi'}\right]_{\mathcal{M}'}=(2-2\Delta)\phi'_{(2\Delta-2)}\  .
\ee
This is an exact result. To write it in terms of $\langle\mathcal{O}(z,\bar{z})\rangle_{\mathbb{C}}$ we need to to write 
$\phi'_{(2\Delta-2)}$ in terms of $\phi_m$'s which can be done perturbatively. Instead, we use the approximate
action $\Tilde{S}_{sub}(\tilde{\Phi},\gamma')_{\mathcal{M}}$ which yields
\be
\langle\mathcal{O'}(w,\bar{w})\rangle_{\mathcal{R}_n}=\lim_{\substack{\epsilon\rightarrow0}}
\left[\frac{1}{\rho^{\Delta/2}\sqrt{\gamma'}}\frac{\delta \tilde{S}_{sub}(\tilde{\Phi},\gamma')}{\delta\tilde{\Phi}}\right]_{\mathcal{M}}=
(2-2\Delta)\tilde{\phi}_{(2\Delta-2)}\  .
\ee
The relation to $\langle\mathcal{O}(z,\bar{z})\rangle_{\mathbb{C}}$ is now obvious after consulting
(\ref{expansions}) 
\be
\langle\mathcal{O'}(w,\bar{w})\rangle_{\mathcal{R}_n}=e^{-\varphi\Delta}\langle\mathcal{O}(z,\bar{z})\rangle_{\mathbb{C}}\ . \label{onepoints}
\ee
Let us summarise what we have done. CFT on $\mathcal{R}_n$ is claimed to be obtained from a regularization of $AdS$ space
on $\mathcal{M}$. Since in the asymptotic region, $\mathcal{M}$ and $\mathcal{M}'$ become equivalent we could use the latter as well.

Since $\mathcal{M}$ is related to $\mathcal{M}_0$ by an exact conformal transformation, fields on the former
are obviously related to those on the latter (see (\ref{expansions})). This makes it easy to relate physical quantities that are extracted from them.
Instead, we have to work out the details of holographic renormalisation method with the regulator $\mathcal{M}$
which is in principle a difficult task.

On the other hand, since $\mathcal{M}'$ has a simple form in the FG coordinates, we already know the form of renormalized
action when using it as a regulator. However, the transformation from $\mathcal{M}'$ to $\mathcal{M}_0$
is a complicated one and reduces to our desired conformal transformation only in the limit.
This in turn makes it difficult to relate physical quantities extracted from $\mathcal{M}'$ to those from $\mathcal{M}_0$.
 
In the above, we have used the form of renormalized action on  $\mathcal{M}'$ to find that on $\mathcal{M}$ approximately.
The final result (\ref{onepoints}) has a suggestive form which guides us to the interpretation and/or justification of
the approximations we have made.
\vspace{0.3cm}

\underline{\emph{Interpretation of the approximations}}
\vspace{0.3cm}

We would now like to see how good an approximation we have used to have replaced $S_{\mathcal{M'}}$ with $\tilde{S}_{\mathcal{M}}$.
A precise answer to this question requires a complete analysis of holographic renormalisation with the regulator $\mathcal{M}$ or,
equivalently, working out a series expansion in $\epsilon$ which relates $\phi'_{m}$ and $\phi_{n}$ based on (\ref{scalars}) and using
the exact transformation (\ref{FGtoP}). For now, we will make use of our answer (\ref{onepoints}) which was obtained approximately.

Remember that upon a conformal transformation $z\rightarrow w$, there are certain operators, primaries, which behave
in a simple form
\be
\mathcal{O}(z,\bar{z})\rightarrow\mathcal{O}'(w,\bar{w})=(\frac{\partial z}{\partial w})^h(\frac{\partial \bar{z}}{\partial \bar{w}})^{\bar{h}}
\mathcal{O}(z,\bar{z})\ ,
\ee
where $h$ and $\bar{h}$ are conformal weights of the primary operator. Comparing this with (\ref{onepoints}) and recalling that
$\Delta=h+\bar{h}$, we have found the conformal properties for a scalar primary operator by holography. We may then
conclude that our approximations are equivalent to assuming the bulk scalar to be dual to a field theory primary.
One can reverse the argument and conclude that if we require our final result to be exact and unaffected by 
corrections we should then impose constraints on the asymptotic expansion of the bulk field and hence 
further relations between the $\phi_n$'s. These restrictions can be interpreted as the bulk definition for a CFT 
primary operator.

Preliminary calculations show that our final result should be corrected by $\phi_{2n}$ terms with $2n<2\Delta-2$ in a generic case.
Recalling that any operator can be expanded in the basis of primaries, we may restate our approximation as  an expansion
of conformal properties of a generic operator in terms of those of the basis operators. The result (\ref{onepoints}) may then be
considered as the term coming from the primary operator in the expansion basis that has the highest dimension.
This problem obviously needs further investigation which is in progress.
\vspace{0.3cm}

\underline{\emph{$n$-point function}}
\vspace{0.3cm}

Recall that in the asymptotic expansion of the scalar field, $\phi_0$ is 
interpreted as the external source for the boundary field theory. We can thus restate our result, (\ref{onepoints}), as following:
in order to find the exact one point function
on $\mathcal{R}_n$, one can use the theory on $\mathbb{C}$ but with a rescaled external source
\be
\phi_0\rightarrow e^{\varphi\Delta}\phi_0\ .\label{source}
\ee
This relation may seem confusing
as $\phi'_0=(e^{\varphi})^{\Delta-2}$ but one should note that the factor of $e^{-2\varphi}$ is cancelled out by
its inverse from the volume element $|dw/dz|^2=e^{2\varphi}$. We expect this result to remain valid when we
turn on interaction terms for the scalar in the bulk and couple the system to gravity. The one point function will
definitely be affected by interactions but what remains unchanged is how a conformal transformation from
$\mathcal{R}_n$ to $\mathbb{C}$ is applied on it. 

Recall that in order to find the $n$-point function 
we should take $n-1$ derivatives from the exact one point function with respect to the external source.
The rescaling of (\ref{source}) will thus lead to
\be
\prod_{i=1}^n\frac{\delta}{\delta\phi_0(z_i,\bar{z}_i)}\rightarrow\prod_{i=1}^ne^{-\varphi(z_i,\bar{z}_i)\Delta}\frac{\delta}{\delta\phi_0(z_i,\bar{z}_i)}\ .\label{recipe}
\ee
\vspace{0.3cm}

\underline{\emph{EE for excitations}}
\vspace{0.3cm}

We can finally use our gravitational results to write (\ref{corrcylinder}), which is defined on a singular cylinder, in the form of
correlators on a smooth cylinder. For the intermediate manifolds, $\mathcal{R}_n^{(w)}$ and $\mathbb{C}$, 
and according to the recipe (\ref{recipe}), we can write
\be
\lim\limits_{\substack{w\rightarrow0\\
w'\rightarrow\infty}}\frac{\langle \displaystyle \prod_{k=0}^{n-1}\CO_k(w,\bar{w})
\CO_k^\dagger(w',\bar{w}') \rangle_{\mathcal{R}_n}}{\langle
\CO(w,\bar{w})\CO^\dagger(w,\bar{w}')\rangle^n_{w}}=
\frac{\displaystyle\prod_{k=0}^{n-1}e^{-[\varphi(z_k,\bar{z}_k)+\varphi(z'_k,\bar{z}'_k)]\Delta}}{\left(\vert(\zeta-1)(\zeta'-1)\vert/l\right)^{2n\Delta}}
\frac{\langle \displaystyle
\prod_{k=0}^{n-1}\CO(z_k,\bar{z}_k)\CO^\dagger(z'_k,\bar{z}'_k)
\rangle_{\MC}}{\langle \CO(\zeta,\bar{\zeta})
\CO^\dagger(\zeta',\bar{\zeta}')\rangle^n_{\MC}}\ ,\ee
where $w=0$ and $w'=\infty$ are the location of in and out insertions on the $w$-plane respectively.
Transformations factors for $\mathcal{R}_n^{(w)}\rightarrow\mathcal{R}_n^{(x)}$ and $\mathbb{C}\rightarrow s$-cylinder, 
are both equal to one and once we plug the values of arguments $z_k,\ \cdots$ in the above expression,
we arrive at our final result 
\begin{equation}
\F_\CO^{(n)}\equiv \frac{tr \rho^n_\CO}{tr \rho^n}=n^{-2n(h+\bar{h})}
\frac{\langle \displaystyle \prod_{k=0}^{n-1}\CO(s_k,\bar{s}_k)
\CO^\dagger(s'_k,\bar{s}'_k) \rangle_{cy}}{\langle
\CO(\xi,\bar{\xi})\CO^\dagger(\xi',\bar{\xi}')\rangle^n_{cy}}\ ,
\ee
where
\be
\xi=\pi\theta\ ,\quad \xi'=-\pi\theta\ ,\quad s_k=\frac{\pi}{n}(\theta+2k)\ ,\quad s_k'=\frac{\pi}{n}(-\theta+2k)\ ,\quad k=0,1,\cdots,n-1\ .
\ee
This is in complete agreement with the field theory result.
$EE$ is now immediately obtained by calculating 
\be
S=-\partial/\partial_n\ tr\rho_{\mathcal{O}}^n\vert_{n=1}\ .
\ee 
Noting that 
$tr\rho_{\mathcal{O}}=\mathcal{F}^{(1)}_{\mathcal{O}}=1$, we find
\be
S_{\mathcal{O}}=S_{GS}-\frac{\partial}{\partial_n}\mathcal{F}^{(n)}_{\mathcal{O}}|_{n=1}\ ,
\ee
where the subscript GS stands for the ground state of the theory.

%%%%%%%%%%%%%%%%%%%%%%%%%%%%%%%%%%%%%%%%%%%%%%%%%%%%%%%%%%%%%%%%%%%%%%%
%%%%%%%%%%%%%%%%%%%%%%%%%%%%%%%%%%%%%%%%%%%%%%%%%%%%%%%%%%%%%%%%%%%%%%%

\section{Discussion}
In this work we have studied the entanglement entropy 
for excited states in a two dimensional
conformal field theory on a circle. 
The entangling area is a single interval and the excited states are
produced by vertex operators. 
We have used holography to address this problem and we were able to reproduce the exact
field theory results for primary excitations.

In a field theory setup, this problem 
boils down to the calculation of certain correlation functions
on a singular  manifold. Since in two dimensions
all metrics are conformally equivalent, one can encode
any singularity in the conformal structure of the space.
Consequently, the problem reduces to a standard calculation on a smooth manifold. 
We simulate the same ideology in the holographic language.

Two dimensional CFT's are conjectured to be described by asymptotically $AdS_3$ spaces.
Such spaces are always locally $AdS$ no matter of whatever modifications we make.
We thus expect that even if the field theory is living on a singular manifold, all the complexities
can be encoded in an appropriately chosen regulator surface in the bulk. 
Regulator surfaces are all conformally equivalent and we choose the conformal class 
that has the previously mentioned singular structure. 

As a point of reference, we also use the usual
smooth regulator surface which is defined at a constant radius of
the Poincare coordinates. 
We show that switching between the two regulators amounts to
performing a conformal transformation on the dual field theory. 
This enables us to make a relation between physical quantities 
that we compute using the two regulation schemes. 
We use perturbative methods to draw such a comparison and conclude that
our approximations are equivalent to assuming certain conformal properties
for the field theory operators.
 
This approach should in principle enable us to find a bulk definition
for conformal properties of the field theory objects, and in particular,
the definition for a field theory primary operator. This subject is
currently under study and we postpone it to a future work.

There are many natural extensions to the present work.
An immediate one is the case with a finite temperature. 
A generalisation to higher dimensions is also of particular interest.
Another interesting direction is the recently discovered
thermodynamic properties for entanglement entropy of
excited states in certain limits \cite{Bhattacharya:2012mi}. The
holographic setup can in principle be used to 
explore such properties in more generality.
%%%%%%%%%%%%%%%%%%%%%%%%%%%%%%%%%%%%%%%%%%%%%%%%%%%%%%%%%%%%%%%

\section*{Acknowledgements}
We would like to thank Mohsen Alishahiha for very useful discussions and also
for a critical reading of the draft.  A.F. would also like to thank Hessamaddin Arfaei for encouragements and support.

%%%%%%%%%%%%%%%%%%%%%%%%%%%%%%%%%%%%%%%%%%%%%%%%%%%%%%%%%%%%%%%

\end{document}